\documentclass[osajnl,twocolumn,showpacs,superscriptaddress,11pt]{revtex4-1} %% use 11pt for Applied Optics

\usepackage{amsmath,amssymb,graphicx}
\usepackage{siunitx}
\usepackage{microtype}

\begin{document}

\title{W-Band Pancharatnam Half Wave Plate Based on Negative Refractive Index Metamaterials}

\author{Imran \surname{Mohamed}}\email{Corresponding author: imohamed@jb.man.ac.uk}
\author{Giampaolo \surname{Pisano}}
\author{Ming Wah \surname{Ng}}
\affiliation{Jodrell Bank Centre for Astrophysics, The Alan Turing Building, School of Physics and Astronomy, The
University of Manchester, Oxford Road, Manchester, M13 9PL, UK}

\begin{abstract}
Electromagnetic metamaterials, made from arrangements of subwavelength sized structures, can be used to 
manipulate radiation. Designing metamaterials that have a positive refractive index along one axis and a 
negative refractive index along the orthogonal axis can result in birefringences, $\Delta n>1$. 
The effect can be used to create wave plates with subwavelength thicknesses. Previous attempts at 
making wave plates in this way have resulted in very narrow usable bandwidths. In this paper, we use the 
Pancharatnam method to increase the usable bandwidth. A combination of Finite Element Method and 
Transmission Line models were used to optimise the final design. Experimental results are compared to the 
modelled data.
\end{abstract}


% REPLACE WITH CORRECT OCIS CODES FOR YOUR ARTICLE
             % NOTE: \ocis{} IS ALIASED TO \pacs{} BUT MUST
             % FORMAT THE TERMS CORRECTLY FOR EACH JOURNAL

\maketitle %% required

\section{Introduction}
Electromagnetic metamaterials, periodically arranged sub-wavelength structures, can be used to interact with radiation 
in ways generally unattainable with natural materials. In the form of metal mesh grids, they have been used since the
1960's by authors such as Ulrich~\cite{ulrich_far-infrared_1967} to create polarisers and filters. The grids properties 
were based on the inductive and capacitive responses that radiation had to the grids geometries. In the past decade 
such grids have been rediscovered primarily for their use in producing negative refractive indices (NRI), and also for 
their ability to be designed with customised electromagnetic properties. This technology finds applications in the 
microwave and terahertz frequency ranges where the availability of suitable dielectric materials that can be used to 
manufacture optical components is very limited.

Most uses for NRI have focused on cases where isotropic behaviour is required e.g.\ in the creation of ``perfect
lenses'' \cite{pendry_negative_2000}. However, polarisation dependent responses can be used to design birefringent 
metamaterials and hence wave plates. Wave plates (or retarders) are optical devices that alter the polarisation 
state of radiation transmitted through them. Half Wave Plates (HWPs) are a particular type of wave plate used to 
arbitrarily rotate the polarisation angle of linearly polarised radiation. They find application when linear polarisation 
modulation is required. In these cases, a HWP rotating at a frequency, $f$, is placed in front of a linearly polarised 
detector. In this setup, the intensity of linearly polarised radiation reaching the detector varies with a frequency of 
$4f$ whilst the intensity of unpolarised radiation is unaffected. This allows the detector to discriminate between 
radiation from polarised and unpolarised sources. 

In this paper, NRI will be used in combination with positive refractive indices (PRI) to create a highly birefringent 
HWP. This idea has been applied in \cite{weis_strongly_2009} where the results were very narrowband. In order in to 
increase the bandwidth in this work the Pancharatnam method 
\cite{pancharatnam_achromatic_1955,pancharatnam_achromatic_1955-1} is employed for first time with NRI based 
wave plates.

\section{Birefringence and Half Wave Plates}
A common way to design wave plates is to use uniaxial birefringent materials. Assuming a Cartesian coordinate system, 
a uniaxial birefringent material has refractive index values such that $n_x \neq n_y = n_z$. Radiation propagating 
through these materials experience different refractive indices depending on the radiation's initial polarisation state and 
propagation direction. This implies that two orthogonal polarisation components can travel at different speeds through 
the material causing a differential phase shift, $\Delta\phi$, so altering the polarisation state from that of the incoming 
radiation. By cutting a slab of a uniaxial birefringent material with its entrance and exit surfaces parallel to the material's 
optic axis, a wave plate is created. For a given frequency, $f$, the value of $\Delta\phi$ produced by a wave plate of 
thickness, $d$, is equal to
\begin{equation}\label{eq:dp}
\Delta\phi = \frac{2\pi f d}{c_0}\Delta n
\end{equation}
where $c_0$ is the free space speed of light and $\Delta n = n_x - n_y$, a quantity known as the birefringence. To
create a HWP, the radiation passing through it must experience a $\Delta\phi$ of \ang{180}.

In the mm-wavelength range, there are not many birefringent materials available for the manufacture of wave plates.
Commonly used  materials include sapphire and quartz, both of which have $\Delta n < 1$ 
\cite{afsar_precision_1987,lamb_miscellaneous_1996}. For a fixed frequency it can be seen from 
equation~\eqref{eq:dp} that the $\Delta\phi$ is proportional to $d$.
So using materials with low birefringence values leads to thicker wave plates with an increased amount of weight and 
absorption compared to thinner wave plates made of the same material. The weight consideration 
is important in cases where large diameters (\SI{>30}{\cm}) are required at cryogenic temperatures.

An alternative to natural materials would be to use the aforementioned metamaterials. By designing them so that they
present a PRI to one polarisation and a NRI to the orthogonal polarisation, a $\Delta n$
greater than unity can be created. This would allow the manufacturing of subwavelength thick wave plates, much thinner 
than those made with birefringent materials. This idea was applied in \cite{weis_strongly_2009} to produce a HWP and a 
QWP. The grids used in \cite{weis_strongly_2009} are similar to the capacitive grids used for mesh filters in 
\cite{ulrich_far-infrared_1967}. To create a NRI the grids were paired up in the radiation's propagation direction.
Resonances in the individual grids result in a negative permittivity, whilst the anti-parallel
currents in the grid pair produces a current loop allowing the production of a negative permeability. When these two
conditions occur in the same frequency range a NRI is produced. The resulting wave plates made in this way however
had very narrow usable bandwidths. This was due to the large gradient of $\Delta\phi$ against $f$ due to the
resonance of the metamaterial required to produce the NRI. If we define the usable bandwidth for a HWP as the region
where $\Delta\phi=\SI{-180\pm3}{\degree}$, the HWP from \cite{weis_strongly_2009} had a fractional bandwidth 
of \SI{\sim0.5}{\percent}.

\begin{figure}[htbp]
\centerline{\includegraphics{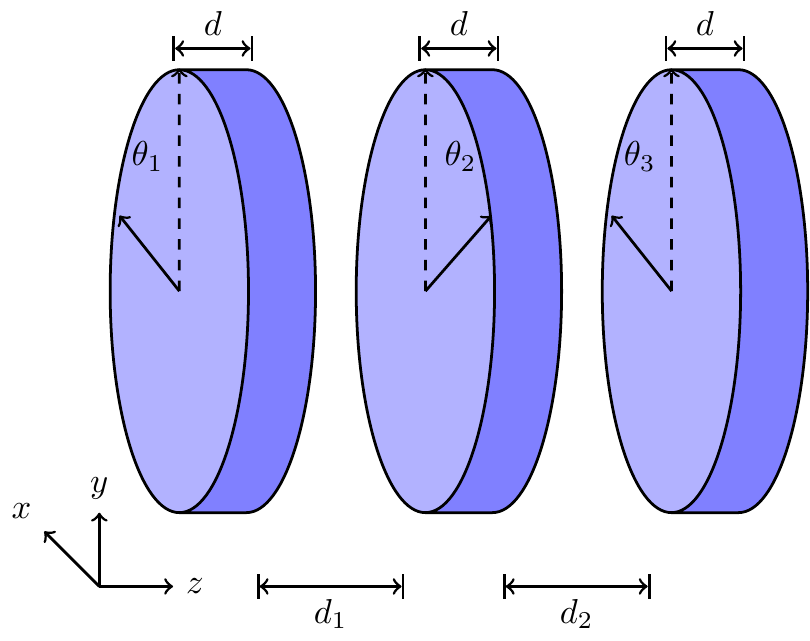}}
\caption{Schematic showing the setup of three wave plates in a Pancharatnam configuration used in this work. The optic axis of each wave plate is rotated by differing angles.}
\label{fig:Panch}
\end{figure}

In this paper we use the Pancharatnam method \cite{pancharatnam_achromatic_1955,pancharatnam_achromatic_1955-1} 
to increase the usable bandwidth. Originally developed for wave plates operating at optical frequencies, the same method 
can be applied at any frequency e.g.\ at millimetre wavelengths \cite{murray_imaging_1997}, as well as with different 
polarisation manipulation technologies e.g.\ waveguide based polarisation rotators \cite{pisano_broadband_2011}. The 
Pancharatnam method requires a number of wave plates to be placed one after the other with their optic axes rotated by 
specific angles. This is similar to the schematic shown in Fig.~\ref{fig:Panch}, although when using wave plates made from 
regular birefringent materials, they are typically in physical contact with one another. A single wave plate of thickness, $d$, 
and birefringence, $\Delta n$, is only able to produce the required $\Delta\phi$ at a single frequency. By using $N$ wave 
plates rotated by different angles, $\theta_1,\theta_2,\ldots,\theta_N$, the required $\Delta\phi$ may be crossed at $N$ 
different frequencies. By optimising the rotation angles of the cascaded set of wave plates, a broadening of the usable 
bandwidth can be achieved. 

\section{Design, Modelling and Optimisation}
Metallic meshes are generally made as a two-dimensional array of repeated structures. These can be 
simulated using commercial electromagnetic Finite Element Method (FEM) solvers such as Ansys HFSS \cite{ansys_hfss_2010} 
by modelling just the smallest repeatable element, the unit cell. By applying periodic boundary conditions to 
the unit cell, an infinite two-dimensional array of cells can be easily simulated without requiring 
large computing resources.

\begin{figure}[htbp]
\centerline{\includegraphics{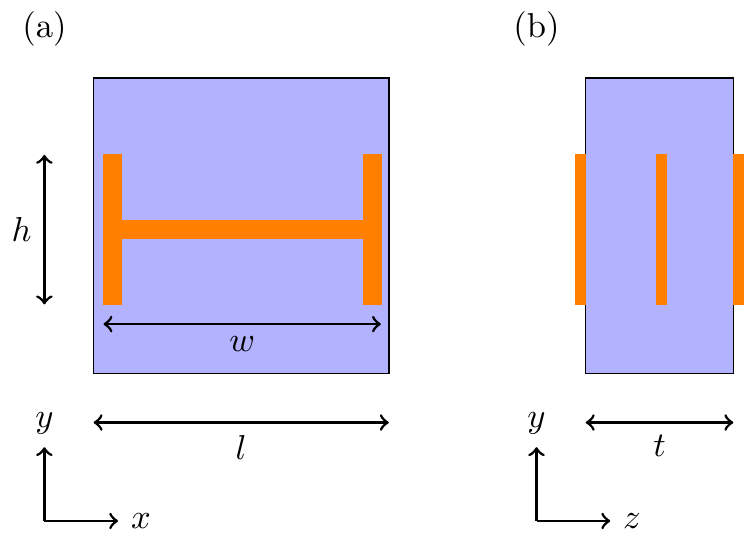}}
\caption{(a)~Front view and (b)~Side view of the dog bone triplet (DBT) unit cell. The copper parts are represented by orange and the polypropylene substrate is coloured light blue. The dimensions are as follows: $h=\SI{300}{\um}$, $w=\SI{556}{\um}$, $l=\SI{591}{\um}$ and $t=\SI{264}{\um}$. The thickness of the copper is \SI{2}{\um}.}
\label{fig:DBT}
\end{figure}

The unit cell design adopted in this work is based on a ``dog bone'' structure (Fig.~\ref{fig:DBT}), previously used in 
\cite{zhou_experimental_2006}. The design was chosen for its birefringent properties, and its ability to produce a NRI 
band only along a single polarisation direction. The final structure consisted of three copper dog bone grids supported 
by a polypropylene ($\varepsilon_r=\num{2.2551}$, $\delta = \num{7e-4}$ \cite{lamb_miscellaneous_1996}) substrate. 
Henceforth this structure will be referred to as a dog bone triplet (DBT). The dimensions of the dog bones and the 
thickness of the substrate were optimised so that a single unit cell would be able to produce $\Delta\phi = \ang{-180}$ 
at \SI{92.5}{\GHz}, the centre of the W-band (\SIrange{75}{110}{\GHz}). Additionally the transmitted intensity along 
the $x$- and $y$-polarisations, $|S_{21}^{x,y}|^2 $ was \num{>0.8} at the same frequency. The final optimised 
dimensions are listed in the caption of Fig.~\ref{fig:DBT}.

Due to the Pancahratnam method requiring a number of wave plates to be cascaded one after the other with 
different arbitrary angles, the available periodic boundary conditions in HFSS could not be used on a cascaded 
set of DBT unit cells. The use of FEM modelling to optimise the Pancharatnam based HWP would have 
required a set of cascaded full sized DBT based HWPs. This would have been computationally prohibitive 
in terms of the CPU time and RAM required.

Our alternative approach is based on a transmission line (TL) model. The use of TL modelling for meshes is well known 
\cite{marcuvitz_waveguide_1965,ulrich_far-infrared_1967} and involves the individual grids being represented by a 
suitable combination of lumped elements e.g.\ capacitors and inductors. Different grid designs can be represented by 
equivalent circuits based on different combinations of lumped elements. Using available analytical equations, the frequency 
dependent behaviour of the grid may be easily computed. When modelling a cascade of grids, each grid can be 
represented by a transmission matrix, such as those described in \cite{pozar_microwave_2012}. The 
matrix elements are still based on the same analytical equations. To calculate the final transmission and reflection of such 
devices, the much less computationally intensive process of matrix multiplication can be carried out. 

However, this method of analytical matrix seeding is unsuitable when very accurate values of the phase are required 
\cite{pisano_broadband_2012}. To achieve accurate phase values, data obtained from FEM simulations of the grids 
were used to ``seed'' the matrix elements, instead of relying on quantities derived analytically. To do this, the 
S-parameters from an FEM simulation are extracted and converted into ABCD transmission matrices 
\cite{pozar_microwave_2012}. The exact formalism of the Pancharatnam method used is based on 
\cite{adachi_analysis_1960,savini_achromatic_2006}, where rotation matrices are used to allow the modelling of the 
wave plates' rotations. A TL model consisting of three DBT HWPs was optimised by varying the rotation angles of the 
individual HWPs and the thickness of the air gaps between them. The optimisation goal was set to maximise the 
frequency range where $\Delta\phi=\ang{-180}$ and the transmitted intensities for both polarisations were equal to 
each other. Using the nomenclature from Fig.~\ref{fig:Panch} the optimised parameters were: 
$\theta_1=\theta_3=\ang{30}$, $\theta_2=\ang{-29}$ and $d_1=d_2=\SI{1.3}{\mm}$.

\section{Manufacture}
The manufacture of the individual dog bone grids was carried out using standard photolithographic techniques. The
process involves evaporating a \SI{2}{\um} layer of copper on to a polypropylene substrate. The copper is
then coated with an even layer of positive photoresist. A photomask consisting of the positive of the grid design
is placed over the sample and the entire system is exposed to ultraviolet (UV) radiation. The sample is then etched
to remove the regions of copper where the photoresist was exposed to the UV radiation, leaving behind
the copper that was protected by the photomask.

To create the individual DBT HWPs, three grids were manually aligned on top of each another with polypropylene spacers
placed between them to provide the necessary separations. The individual grids and layers of polypropylene were then
hot pressed together inside a vacuum oven to form a single structure. To provide support to the HWPs as well as
provide the required thickness of the air gaps between them, the HWPs were mounted on \SI{1.3}{\mm} thick aluminium rings.
Transmission measurements of the individual HWPs were then taken to test their performance before they were combined
into the full Pancharatnam HWP structure.

The Pancharatnam HWP was then created by stacking the individual ring mounted HWPs on top of each other. To ensure the
HWPs were rotated by the correct amount the required rotation angles were marked on to the circumference of the rings
and the markings were used to aid the aligning.

\section{Experimental Setup}
\begin{figure}[htbp]
\centerline{\includegraphics{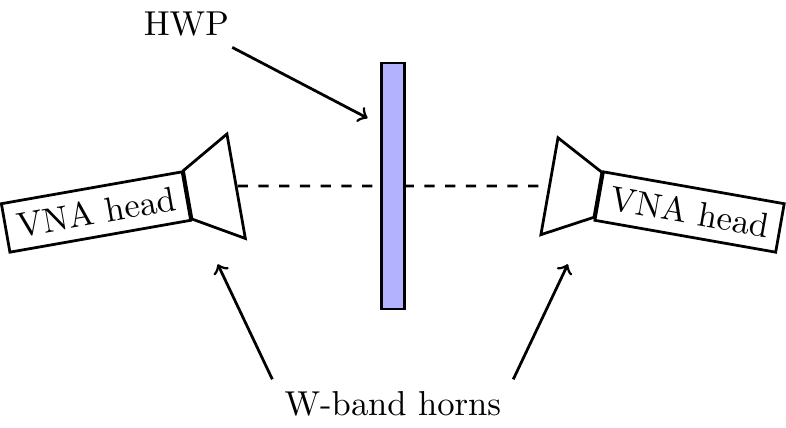}}
\caption{Schematic of experimental setup used for the transmission readings. The VNA heads are aligned off-axis to one another whilst keeping the line that joins the centres of the two heads perpendicular to the HWP.}
\label{fig:Setup}
\end{figure}

Normal incidence transmission measurements were taken using the experimental setup shown in Fig.~\ref{fig:Setup}. 
The Pancharatnam HWP is held in a specially 
designed holder between the two W-band heads of a Rhode \& Schwarz ZVA40 Vector Network Analyser (VNA). 
The heads are placed off-axis to one another but at the same height and in such a way that the line connecting the 
two heads bisect the HWP perpendicularly. Having the heads off-axis allows the reduction of unwanted standing 
waves between the HWP and the heads. Although the HWP is no longer illuminated using the central portion of the 
horn beams, the polarisation vectors of the emitting and receiving heads are still aligned. This is crucial for 
polarisation measurements where cross polarisation leakages needs to be minimised. The requirement that the line 
connecting the horns perpendicularly bisect the HWP means that it is illuminated by a nearly planar wave front, 
closely matching the conditions that were modelled. Eccosorb was appropriately placed around the setup to further 
reduce unwanted reflections.

Due to the existence of a NRI along the $x$-axis, $x$-polarised radiation incident at non-normal angles would be 
refracted in the opposite manner within the wave plate than if it had a PRI. For a single DBT this behaviour could be 
detected by measuring the lateral shift of the transmitted beam \cite{alici_direct_2009}. However in our 
experimental setup, although the VNA heads were positioned off-axis, the radiation incident on the plate was at 
almost normal incidence and the effect was neither expected nor observed. In the case of the full Pancharatnam 
HWP illuminated with off-axis radiation, the behaviour would be even more complicated, both due to the DBTs optic 
axes being rotated with respect to one another as well as the polarisation dependent behaviour observed with 
non-normal incidence on metal mesh grids \cite{pisano_polarisation_2006}. 

\section{Data Analysis}
\subsection{Single HWP Performance}
\begin{figure}[htbp]
\centerline{\includegraphics{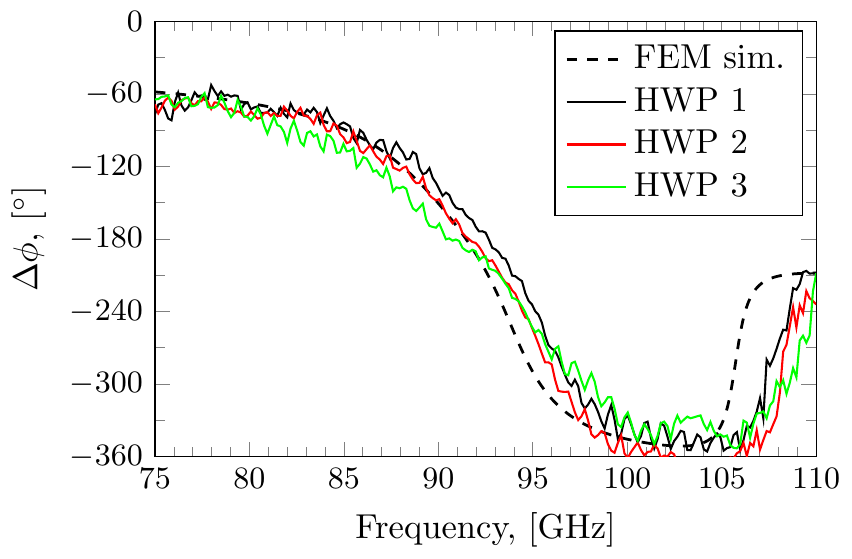}}
\caption{The FEM simulated (dashed) and VNA measured (solid) transmitted phase difference of the individual DBT HWPs.}
\label{fig:DBT-dp-123}
\end{figure}

A FEM simulated frequency sweep of the optimised DBT grid showed that a single HWP made from a single DBT 
grid was able to meet the set requirements. $\Delta\phi = \SI{-180\pm3}{\degree}$ between \SI{91.3}{\GHz} and 
\SI{93.6}{\GHz}, providing a usable fractional bandwidth of \SI{0.3}{\percent} (Fig.~\ref{fig:DBT-dp-123}, dashed line). 
This narrow bandwidth is caused by the large magnitude of $\Delta\phi$'s gradient when it crosses 
\ang{-180}. The steepness is a result of the resonance required to produce a NRI region. VNA readings 
were taken of the individual HWPs to test their performance before combining them to form the Pancharatnam 
HWP and the results are shown in Fig.~\ref{fig:DBT-dp-123} as solid lines. All three manufactured HWPs show agreement 
with the FEM simulation data of Fig.~\ref{fig:DBT-dp-123} (dashed line). HWPs 1 and 2 show the best agreement, only 
drifting from the FEM data above \SI{\sim90}{\GHz} where the measured data becomes blue-shifted by 
\SI{\sim2}{\GHz}. HWP~3 differs slightly more than the first two HWPs with its measured $\Delta\phi$ showing a 
shallower gradient causing its $\Delta\phi$ to be lower than the expected value between \SIrange{75}{92}{\GHz} 
and greater than the expected value above \SI{92}{\GHz}. 

\begin{figure}[htbp]
\centerline{\includegraphics{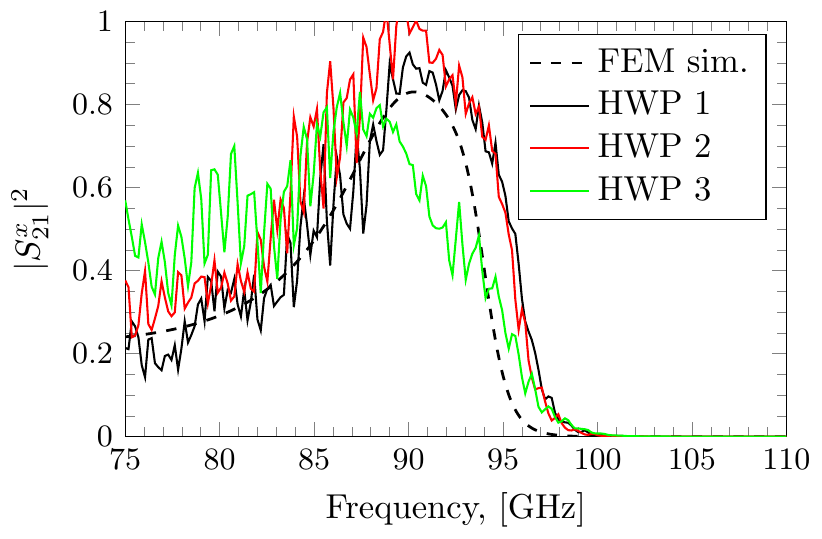}}
\caption{The FEM simulated (dashed) and VNA measured (solid) transmitted intensity along the $x$-axis of the individual DBT HWPs.}
\label{fig:DBT-Ix-123}
\end{figure}

\begin{figure}[htbp]
\centerline{\includegraphics{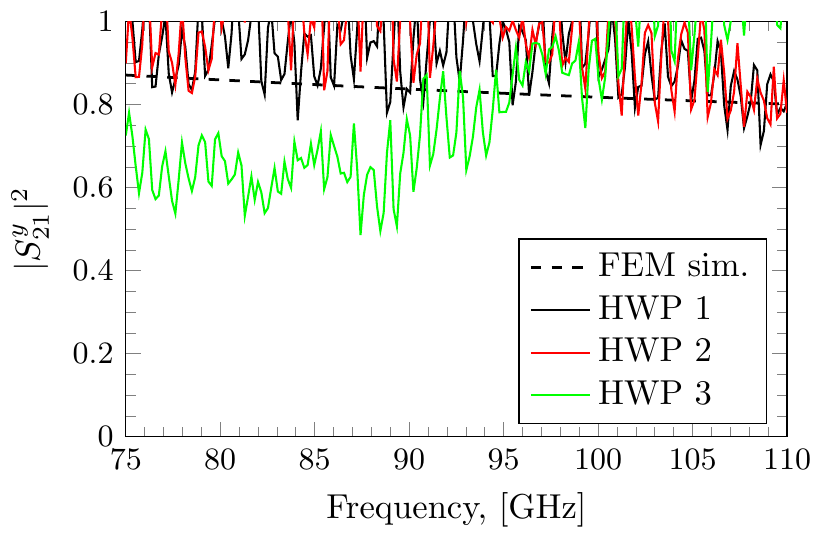}}
\caption{The FEM simulated (dashed) and VNA measured (solid) transmitted intensity along the $y$-axis of the individual DBT HWPs.}
\label{fig:DBT-Iy-123}
\end{figure}

The simulated transmitted intensities along the $x$- and $y$-polarisations, $|S_{21}^{x}|^2$ and $|S_{21}^{y}|^2$, 
are shown as dashed lines in Fig.~\ref{fig:DBT-Ix-123} and Fig.~\ref{fig:DBT-Iy-123}. Within the defined usable bandwidth, 
the mean $|S_{21}|^2$ are \num{0.81} for $x$-polarised radiation and \num{0.83} for $y$-polarised radiation. Whilst 
$|S_{21}^{x}|^2$ shows a resonance, peaking at \SI{90.3}{\GHz}, $|S_{21}^{y}|^2$ remains unaffected by the DBT, 
instead showing a slow decrease in intensity, but remaining above \num{0.8} across the entire W-band. The measured 
$|S_{21}^{x}|^2$ and $|S_{21}^{y}|^2$ of the individual HWPs are shown in Fig.~\ref{fig:DBT-Ix-123} and 
Fig.~\ref{fig:DBT-Iy-123} as solid lines. Like the $\Delta\phi$ data we see that HWPs 1 and 2 show good agreement with the 
expected data. However HWP~3 shows deviations for both $x$- and $y$-polarisations. 

The cause for the observed deviations in $\Delta\phi$, $|S_{21}^{x}|^2$ and $|S_{21}^{y}|^2$ from the expected 
behaviour is misalignment between the grids that make up the individual HWPs. These misalignments can be translational 
and rotational, with only the former being easily simulated with the single unit cell and periodic boundary condition setup 
in HFSS. 

\begin{figure}[htbp]
\centerline{\includegraphics{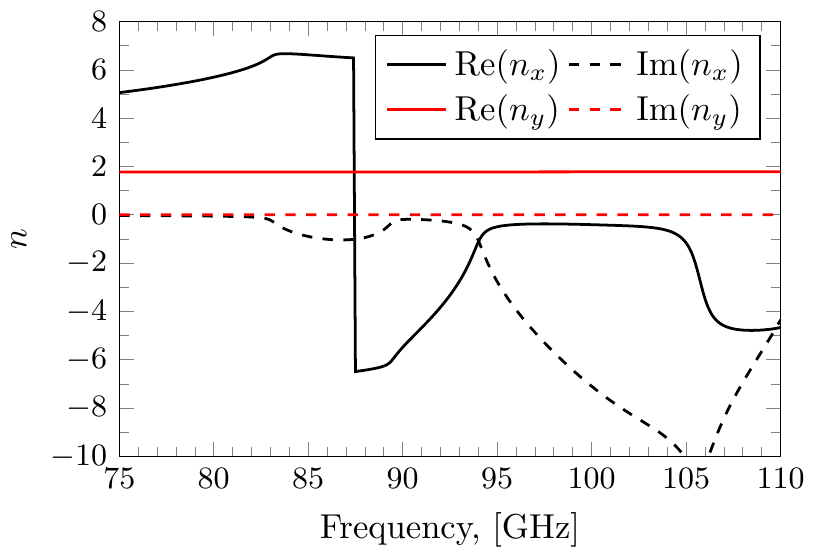}}
\caption{The refractive indices, $n_x$ and $n_y$, of the $x$- and $y$-polarised radiation calculated from FEM simulation data of the DBT.}
\label{fig:DBT-n}
\end{figure}

The simulated refractive indices, $n_x$ and $n_y$, are shown in Fig.~\ref{fig:DBT-n}. In the $x$-axis, the HWP is able 
to produce a NRI band above \SI{87.5}{\GHz} whilst the refractive index along the $y$-axis remains constant. The values were 
obtained using the $S_{11}$ and $S_{21}$ data and applying a parameter extraction method \cite{chen_robust_2004}. 
The thickness used in the parameter extraction calculation was the physical thickness of the DBT grid, 
\SI{268}{\um}, the sum of the substrate thickness, $t$, and the thicknesses of the copper on the outer 
surfaces (Fig.~\ref{fig:DBT}).

\subsection{Pancharatnam HWP Performance}
\begin{figure}[htbp]
\centerline{\includegraphics{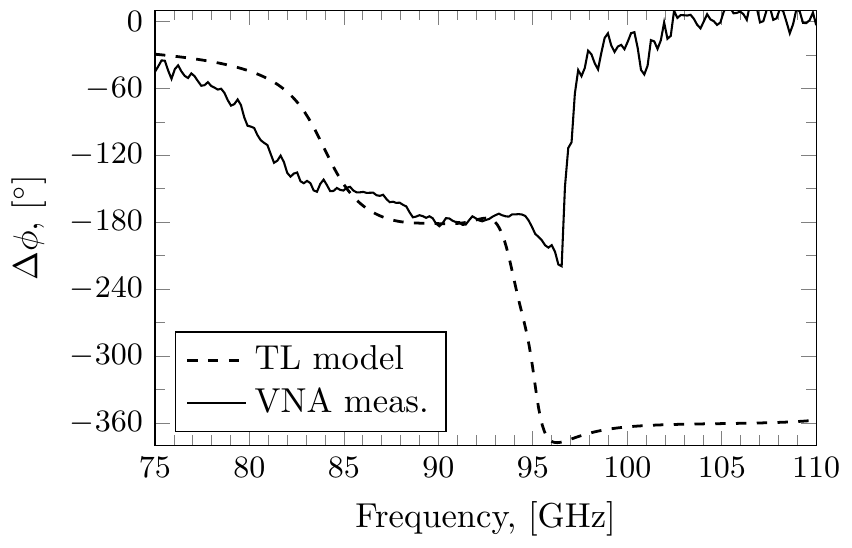}}
\caption{The TL modelled (dashed) and VNA measured (solid) transmitted phase difference of the Pancharatnam based HWP.}
\label{fig:HWP-dp}
\end{figure}

Using the TL based model with the optimised parameters, the expected $\Delta\phi$ is shown as the dashed line 
in Fig.~\ref{fig:HWP-dp}. Comparing this to the behaviour of the single HWPs it can be seen that there is a flattening 
of $\Delta\phi$ between \SIrange{87.4}{92.2}{\GHz}. This means the usable fractional bandwidth has been increased by 
almost 18~times to \SI{5.3}{\percent}. This demonstrates that even if the individual HWPs have very high gradients of $\Delta\phi$, 
the Pancharatnam method can still produce a flat region in $\Delta\phi$. Broader bandwidths can be attained by using 
more HWPs.

\begin{figure}[htbp]
\centerline{\includegraphics{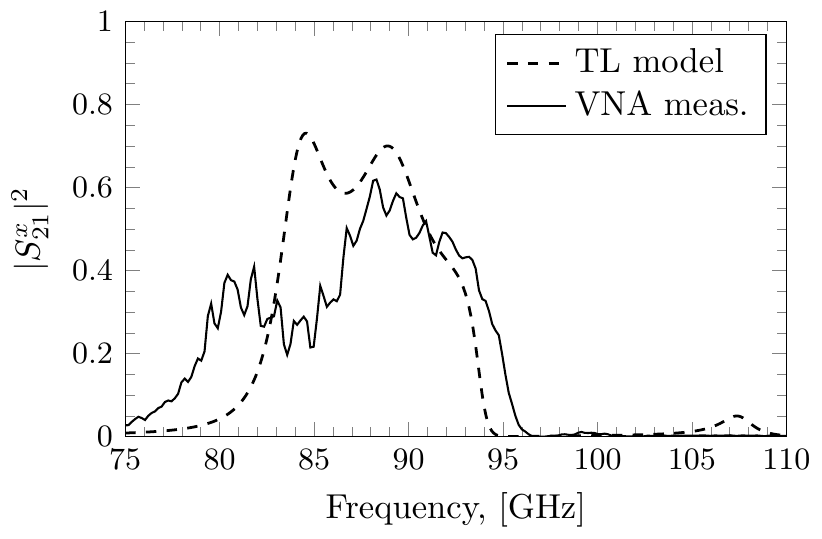}}
\caption{The TL modelled (dashed) and VNA measured (solid) transmitted intensity of the Pancharatnam based HWP along its $x$-axis.}
\label{fig:HWP-Ix}
\end{figure}

The measured readings of $\Delta\phi$, shown as the solid line in Fig.~\ref{fig:HWP-dp}, shows some agreement with 
the expected data from the TL model between \SIlist{75;93}{\GHz}. The important result to note is a plateau of 
$\Delta\phi=\SI{-180\pm3}{\degree}$ is observed albeit only between \SIrange{88.9}{92.7}{\GHz}, a fractional bandwidth 
of \SI{3.1}{\percent}. Above \SI{96.7}{\GHz}, $\Delta\phi$ appears to deviate greatly from the TL model, but it should noted that in 
this region $|S_{21}^{x}|^{2}$ goes to zero (Fig.~\ref{fig:HWP-Ix}), meaning accurate measurements of 
arg($S_{21}^x$) becomes difficult to obtain.

\begin{figure}[htbp]
\centerline{\includegraphics{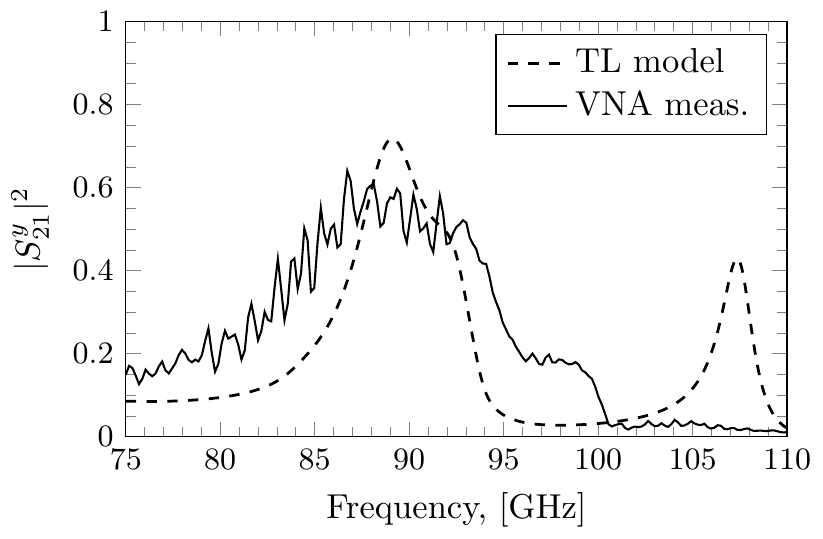}}
\caption{The TL modelled (dashed) and VNA measured (solid) transmitted intensity of the Pancharatnam based HWP along its $y$-axis.}
\label{fig:HWP-Iy}
\end{figure}

The TL modelled values of $|S_{21}^{x}|^2$ and $|S_{21}^{y}|^2$ are shown as dashed lines in Fig.~\ref{fig:HWP-Ix} 
and Fig.~\ref{fig:HWP-Iy} respectively. The expected $|S_{21}|^2$ for the $x$- and $y$-polarisations is reduced compared 
to the performance of a single HWP. In the usable band, the mean values of $|S_{21}^{x}|^2$ and $|S_{21}^{y}|^2$ were 
\num{0.59} and \num{0.60} respectively, with the peak values reaching \num{0.70} and \num{0.72} respectively around \SI{\sim89}{\GHz}. The transmitted 
intensities of the two polarisations also show minimal dichroism within this band with the two values showing close overlap. 

The measured readings of $|S_{21}^{x}|^2$ and $|S_{21}^{y}|^2$ are shown as solid lines in Fig.~\ref{fig:HWP-Ix} 
and Fig.~\ref{fig:HWP-Iy} respectively. The measured data shows some agreement with the expected TL modelled 
data, being lower in intensity and flatter in profile, with the peak intensities in the usable band for both polarisations 
being around \num{0.60}.

The discrepancies shown in the measured $\Delta\phi$, $|S_{21}^{x}|^{2}$ and $|S_{21}^{y}|^{2}$ are due to the deviations 
from the expected behaviour of the individual HWPs, especially that of HWP~3. Remarkably however, despite 
these deviations, flattening of $\Delta\phi$ is still observed, demonstrating that the HWPs made via the 
Pancharatnam method show some resilience to manufacturing errors.

\section{Conclusions}
In this paper, a highly birefringent metamaterial was created that was capable of producing a NRI and a PRI along 
orthogonal polarisation axes. This was used to create a HWP with a narrow usable fractional bandwidth defined 
as $\Delta\phi=\SI{-180\pm3}{\degree}$ of \SI{0.3}{\percent}. Three of these HWPs were cascaded according to the Pancharatnam 
method to increase the usable bandwidth. 
TL based modelling of the Pancharatnam setup was able to increase this substantially, bringing the fractional bandwidth to \SI{5.3}{\percent}. 
Experimentally measured data taken with the manufactured Pancharatnam HWP 
showed a flattening of \SI{3.1}{\percent} in bandwidth between \SIrange{88.9}{92.7}{\GHz}. The reduced performance was due 
to the misalignment of the grids that made up the individual HWPs. 

This work verifies the ability of the Pancharatnam method to create a broadening of the region where 
$\Delta\phi$ is flat even when the initial $\Delta\phi$ gradients of the constituent wave plates is very steep. 
Broader bandwidths would be achievable by cascading a greater number of HWPs, but this would come at the 
price of lower transmitted intensities. 

\begin{acknowledgements}
The first author was funded by a studentship provided by the Science and Technology Facilities Council (STFC).
\end{acknowledgements}

\end{document}